\documentclass[
final,
5p,
times,
twocolumn,
amssymb
]{elsarticle}

\usepackage{bm}
\usepackage{dcolumn}
\usepackage{graphicx}
\usepackage[utf8]{inputenc}
\usepackage[english]{babel}
\usepackage{amssymb}
\usepackage{amsmath}
\usepackage{mathtools, cuted}

\biboptions{sort&compress}

\newenvironment{equationc}{\begin{equation}}{,\end{equation}\ignorespacesafterend}
\newenvironment{equationp}{\begin{equation}}{.\end{equation}\ignorespacesafterend}

\begin{document}

\begin{frontmatter}
\title{A Boson exchange approach for Helium Burning Stars}


\author{ T. Depastas$^{a,*}$ and A. Bonasera$^{a,b}$ }


\address{ $^{a}$ Cyclotron Institute, Texas A\&M University,
                     College Station, Texas, USA }
\address{ $^{b}$ Laboratori Nazionali del Sud, INFN, Catania 95123, Italy }

\address{ $^{*}$ Corresponding author. Email:  tdepastas@tamu.edu}

\begin{abstract}
Helium burning plays a key role in the hierarchy of stellar nucleosynthesis and evolution, as the archetype of a bosonic 3-body system. Here, we examine the kinetic and nuclear aspects of the $3\alpha$ reaction, under the auspices of the Thomas-Efimov theorem. Due to the $92.08$ keV ground state of $^8$Be, multiple $\alpha$-cluster resonances appear especially at the lowest temperature (Thomas State), while with increasing temperature the system is dominated by the Hoyle/Efimov state. We extend our previous methodology for the sequential channel to describe the direct mechanism, that results in an equilateral geometry similar to the Thomas state. This is accomplished by developing a general approach to the N-body scattering, via successive particle-exchanging 2-body collisions, which eliminates long range Coulomb complications, in a similar manner to the Thomas-Efimov mechanism. We furthermore, discuss the $e^+e^-$ decay of the compound state with a perturbation based methodology. Due to the symmetry of the system and the resulting reaction rates, we favor the $E0$ decay scheme over the $E2$. The $E0$ reaction rates obey all the available astrophysical and nuclear constraints and are compared to theoretical data from the literature, whose limitations are discussed. Our extended methodology provides a physically sound description of the debated low temperature region.
\end{abstract}


\end{frontmatter}

\begin{figure*}[!ht]                                        
\centering
\includegraphics[height=4.0 cm]
{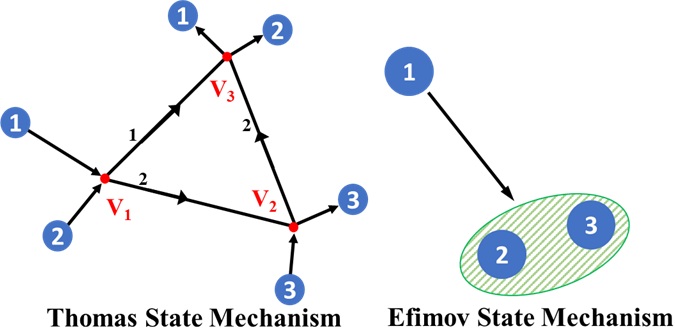} 
\caption{(Color online) Diagrammatic representations of the Thomas (left panel) and Efimov (right panel) mechanisms. The vertices ($V_{i=1,2,3})$ and intermediate particle trajectories are denoted for the former.}
\label{Fig1}
\end{figure*}

\begin{figure}[!ht]                                        
\centering
\includegraphics[height=6.0 cm]
{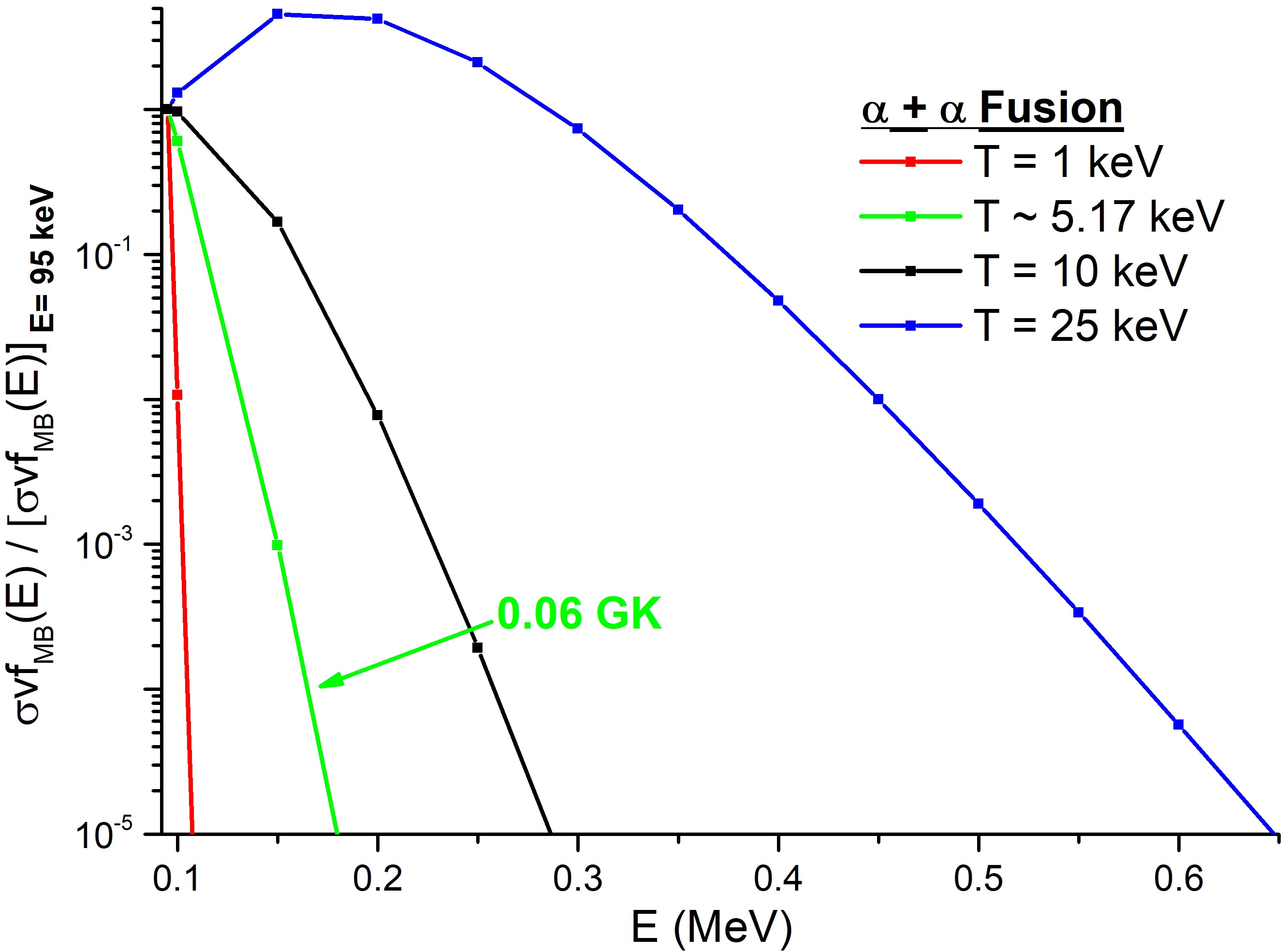} 
\caption{(Color online) The integrad quantity $\sigma_{\alpha\alpha}(E)f_{MB}(E)$ normalized by its value at $95$ keV as a function of the $\alpha-\alpha$ energy for different temperatures, according to the key. The data are obtained via the H$\alpha$C model, as described in Ref \cite{Depastas2024EPJ}.}
\label{Fig5}
\end{figure}

\indent Helium (He-) burning or 3$\alpha$ process is the cornerstone of stellar evolution and nucleosynthesis. As the connecting link between the Red Giant (RGB), Horizontal (HB) and Asymptotic Giant (AGB) Branches \cite{Hayashi1962}, it bridges the $A=5-8$ stable isotope gap \cite{Iliadis2007} and forms the main source of the life-determining element of carbon in the universe. First being conceptualized in the early 1950's by Saltpeter \cite{Saltpeter1952} and experimentally observed by Dunbar et al. and Fowler et al. \cite{DunbarPR1953,CookPR1957}, the 3$\alpha$ reaction remains a central topic of discussion in the literature for more than seven decades. Currently, the process is understood to occur through two competing mechanisms, the sequential-$(2+1)\alpha$ or resonant and the direct-$3\alpha$ or non-resonant.\\
\indent The resonant channel, which dominates in the high temperature region ($T\gtrsim10^8$ K) \cite{Hayashi1962} considers the two step reaction of 2-$\alpha$ particles forming an excited $^8$Be nucleus and the subsequent capture of another $\alpha$ towards an excited $^{12}$C system. The necessary condition for this mechanism to occur is the compound nucleus $\gamma$-decay to the ground state, which is ensured by the $0^+_2$ resonance predicted by Hoyle at $E_H=7.654$ MeV \cite{Hoyle1954}. The non-resonant channel, by contrast, assumes the direct fusion of 3-$\alpha$ particles without the presence of any resonances in the compound system. As such, this mechanism is thought to be dominant at low temperatures, in the $10^7$-$10^8$ K region \cite{Nomoto1985}, while its contribution near $E_H$ is estimated at $\lesssim0.5\%$ that of the sequential channel \cite{Cardella2022}.\\
\indent Despite their differences, the two mechanisms are essentially two aspects of the same phenomenon, the Thomas-Efimov effect. The Thomas theorem \cite{Thomas1935}, states that a loosely bound quantum 2-body system gives a tightly bound 3-body system, whose binding energy tends to infinity in the limit of a force with zero range. Extending this idea, the Efimov effect \cite{Efimov1970} predicts an infinite geometric series of excited levels in the 3-body system, when the scattering length becomes significantly larger than the range. The physical intuition behind the combined phenomenon is the existence of the 3-body system in a state of mutual 2-body resonances, akin to a de-localized particle exchange between two others \cite{Naidon2017}. Following the interpretation of Zheng and Bonasera \cite{ZhengEfimov2020}, the Thomas State (TS) corresponds to the state of the system with an equilateral geometry which decays into 3 monomers, while the Efimov State (ES) is an excitation that decays into a dimer-monomer configuration. The unification of the sequential and direct channels is then clear, the former describes the fusion of an ES and the latter of a state akin to the TS, both of them followed by de-excitation to the ground state of $^{12}$C (which is not a TS \cite{ZhengEfimov2020}).\\
\indent Adopting the unified Thomas-Efimov understanding of He-burning, in this work we strive to extend our framework for the sequential channel \cite{Depastas2024EPJ,Depastas2024Plb} to the direct. To briefly summarize our sequential approach, in Ref. \cite{Depastas2024Plb} we consider the Imaginary Time \cite{BonaseraITM1994,BonaseraITM2000,Depastas2023} fusion of the compound $^8$Be$^*$ and $^{12}$C$^*$ nuclei via the Hybrid $\alpha$-Cluster (H$\alpha$C) \cite{ZhengHac2021} and Neck (NM) \cite{BonaseraNM1984} models and their $\alpha$ and $\gamma$ decays. The two decay widths are taken proportional to each other and inversely proportional to the tunneling time $\tau$, which includes a non-constant proportionality factor to the penetrability $T_0$. Along with the choice of the $^8$Be energy $E_{^8Be}=92.08$ keV as the low integral limit in the reaction rate (which is also discussed later in the text), our approach provides corrections to the by now, standard methodology of the NACRE collaboration \cite{Angulo1999}, leading in a $\sim35$ orders of magnitude rate reduction at low temperatures \cite{Depastas2024Plb}.\\
\indent Our approach to the direct $3\alpha$ channel is presented as follows. First, we devise Thomas-Efimov based formulas to obtain the N-body fusion probability, following closely the ideas of Ref.s \cite{Bonasera3Body1991,BonaseraMeanFreePath1992,BonaseraBoltzmann1994}. Afterwards, we describe the $(2+1)\alpha$ and $3\alpha$ kinetics, followed by an investigation of the possible ES decay paths. Then, we present our resulting reaction rates and compare them to the literature. \\
\indent The probability of a simultaneous scattering of N identical particles $\delta \Pi_N$ in a time interval $\delta t$ is defined via the corresponding mean free path $\lambda_N$ as \cite{Bonasera3Body1991}:
\begin{equationc}
\delta \Pi_N \equiv \frac{1}{\lambda_N}v\delta t
\label{e1}
\end{equationc}
where $v\leq v_{ij},\forall i,j \leq N$ is the relative velocity of the pair that determines the probability rate. In the assumption of only 2-body forces, the N-body interaction may be considered as a series of virtual 2-body collisions that take place in much faster time scales that the typical translational motion \cite{BonaseraBoltzmann1994}. Although it is well known that an explicit repulsive 3-body force is necessary for the stability of a macroscopical Fermionic system \cite{BonaseraBoltzmann1994,VautherinBrinkPRC1972}, the density of the stellar medium is much lower than the typical nuclear matter value of $\rho_0\sim0.16$ fm$^{-3}$, so this contribution can be neglected. In the limit of molecular chaos and a near homogeneous medium\cite{BonaseraMeanFreePath1992}, each successive collision between the $N^{th}$ and $(N-1)^{th}$ particles happens independently, if the former is in the effective interaction volume $\widetilde{V}$ of the latter. The probability $\widetilde{\Pi}$ of this event to occur is \cite{BonaseraBoltzmann1994}:
\begin{equationp}
\widetilde{\Pi}=\rho\widetilde{V}=\rho\frac{4\pi}{3}\left(\frac{\sigma}{\pi}\right)^{3/2}=\frac{4}{3\sqrt{\pi}}\rho\sigma^{3/2}
\label{e2}
\end{equationp}
\indent The N-scattering probability of Eq. \ref{e1} may be then calculated recursively for each collision:
\begin{equationc}
\delta \Pi_N = \delta \Pi_{N-1}\widetilde{\Pi} = ... = \delta \Pi_{2}\widetilde{\Pi}^{N-2}=\frac{v\delta t}{\lambda_2}\left(\frac{4}{3\sqrt{\pi}}\rho\sigma^{3/2}\right)^{N-2}
\label{e3}
\end{equationc}
with the 2-body mean free path given by $\lambda_2=1/\rho\sigma$. The final relation for the N-body scattering probability is then given by:
\begin{equationp}
\delta \Pi_N = \left(\frac{4}{3\sqrt{\pi}}\right)^{N-2}\rho^{N-1}\sigma^{3N/2-2}v\delta t
\label{e4}
\end{equationp}
\indent To intuitively understand this process, we consider the diagrammatic representation of Fig. \ref{Fig1} for $N=3$, which is similar to the one from Ref. \cite{BonaseraBoltzmann1994}. Unlike the sequential scattering (right panel), the direct 3-body scattering (left panel) consists of three 2-body collisions and is irreducible, since particles (1) and (2) collide twice. This traps the system in a cyclic mutual resonance, pictured as the exchange of particle (2), between (1) and (3). This is exactly the Thomas-Efimov mechanism and is considered a fitting description of near-homogeneous stellar cores, where the $\alpha$-particles exist in a sea of $^8$Be resonances, keeping their density almost constant \cite{Nomoto1985}. The existence of such mutual $^8$Be compound systems is strong enough for fusions to take place and thus, there is no requirement of a 3-body resonant state. This is the beauty of the Thomas-Efimov approach and along with its ease of continuum calculations, especially for charged particles which are known to present convergence issues \cite{Suno2016} and its generalized N-body character makes it advantageous for such studies. The discussion of the implications of $(N> 3)$-body collisions are nevertheless reserved for a future paper.\\
Building on the microscopic scattering N-body probability, we can describe the kinematics of He-burning. The reaction rate per unit volume $R_N$ for a system composed of $\nu_1$ particles of type $1$, $\nu_2$ particles of type $2$, ..., is defined as \cite{Hayashi1962}:
\begin{equationc}
R_N \equiv \frac{\rho_1^{\nu_1}\rho_2^{\nu_2}...}{\nu_1!\nu_2!...}\widetilde{R}_N
\label{e5}
\end{equationc}
with $\widetilde{R}_N$ the reduced reaction rate, representing the rate per N-tuple of particles. For identical particles and one-step processes, the rate can be written as:
\begin{equationc}
R_N =\frac{\rho}{N!}\langle\frac{\delta \Pi_N}{\delta t}\rangle=\frac{\rho^N}{N!}\left(\frac{4}{3\sqrt{\pi}}\right)^{N-2}\langle\sigma^{3N/2-2}v\rangle\equiv \frac{\rho^N}{N!}\widetilde{R}_N
\label{e6}
\end{equationc}
where the $N!$ factor takes into account the different particle permutations and the symbol $\langle ...\rangle$ implies averaging with the Maxwellian of relative motion \cite{Angulo1999}, i.e.,
\begin{equationc}
\langle ...\rangle  = A_{MB}\int_0^{\infty}...Ee^{-E/kT}dE\equiv A_{MB}\int_0^{\infty}...f_{MB}(E)dE
\label{e6a}
\end{equationc}
with $A_{MB}=\sqrt{\frac{8}{\mu\pi\left(kT\right)^3}}$. We note that bosonic correlations of the form $\prod_i(1+f_i)$ (with $f_i$ the 1-body distribution function) that amplify the 2-body interactions are neglected, due to the low density of a He-burning star, in comparison with the typical nuclear value. With these definitions, the reduced reaction rate ($\widetilde{R}_3\rightarrow\langle3\alpha\rangle_X$) for the $3\alpha$ channel is given by:
\begin{equationc}
\langle3\alpha\rangle_X=\frac{4}{3\sqrt{\pi}}\langle \sigma_{\alpha\alpha}^{5/2}v P_X\rangle
\label{e7}
\end{equationc}
with $\sigma_{\alpha\alpha}$ the $\alpha+\alpha \rightarrow ^8$Be fusion cross section, which is taken from the H$\alpha$C results of Ref.s \cite{Depastas2024EPJ,Depastas2024Plb} and $P_X$ the decay probability via an exit path ``$X=e^+e^-,\gamma\gamma$", which is discussed later in the text. The reduced rate of the $\left(2+1\right)\alpha$ channel is, by contrast, obtained under the assumption of a Steady State Approximation (SSA) in the $\alpha+\alpha \rightleftharpoons$ $^8$Be equilibrium and the rate determining $\alpha$+$^8$Be $\rightarrow$ $^{12}$C + $\gamma\gamma$ step \cite{Nomoto1985}. In total, the sequential reduced rate ($\widetilde{R}_3\rightarrow\langle\left(2+1\right)\alpha\rangle_{\gamma\gamma}$) in the compound nucleus approach, is calculated via the relation \cite{Nomoto1985,Angulo1999}:
\begin{equationc}
\langle\left(2+1\right)\alpha\rangle_{\gamma\gamma}=3\langle \sigma_{\alpha\alpha}v\frac{\hbar}{\Gamma_{\alpha\alpha}} \langle \sigma_{\alpha^8Be} P_{\gamma\gamma}v\rangle \rangle
\label{e8}
\end{equationc}
with $\sigma_{\alpha^8Be}$ being the $\alpha$+$^8$Be$\rightarrow$ $^{12}$C+$\gamma\gamma$ fusion cross section, $\Gamma_{\alpha\alpha}$ the decay width of the first step, while the $v$ symbols in the outer and inner braces correspond to the relative velocities of $\alpha-\alpha$ and $\alpha-^8$Be, respectively. As shown in Ref \cite{Depastas2024Plb}, the reduced rates for the two steps are almost independent for energies below the Coulomb barrier and thus, Eq. \ref{e8} may be approximated as:
\begin{equationp}
\langle\left(2+1\right)\alpha\rangle_{\gamma\gamma}\approx3\langle \sigma_{\alpha\alpha}v\frac{\hbar}{\Gamma_{\alpha\alpha}}\rangle\langle \sigma_{\alpha^8Be} P_{\gamma\gamma}v\rangle
\label{e9}
\end{equationp}
\indent We stress that, similarly to our sequential study \cite{Depastas2024EPJ,Depastas2024Plb}, the lower limit of integration of the $\alpha-\alpha$ system in Eq. \ref{e7}-\ref{e9} is $E_{\alpha\alpha}\ge E_{^8Be}> 0$. In order for the endothermic $\alpha+\alpha \rightarrow ^8$Be, $Q_{\alpha\alpha}=-E_{^8Be}$ reaction to proceed into fusion, it is required for the excitation energy of the compound nucleus to be at least zero, i.e., $E^*_{^8Be}=E_{\alpha\alpha}+Q_{\alpha\alpha}=E_{\alpha\alpha}-E_{^8Be}\ge 0$. This is a statement of conservation of energy and can be violated according to the Heisenberg principle, but only for short time scales, much shorter than the already short life-time of $^8$Be \cite{Depastas2024EPJ}. Consequently, for energies less than the $^8$Be resonance, the compound system would decay long before the third $\alpha$-particle is captured and the sequential $3\alpha$ process would not proceed.\\
\indent In terms of a purely quantum argument, the previous intuitions may be understood as an application of the variational principle to the shallow $\alpha-\alpha$ potential, in which $^8$Be is considered a quasi-bound state. The energy of such a localized state does not belong in the eigen-spectrum of the Hamiltonian, but is constrained from below by the Hamiltonian matrix elements in an exactly-solvable basis, under the assumption of negligible mixing with the excited $\alpha-\alpha$ states. All this discussion is encapsulated by the condition of the $\alpha-\alpha$ fusion cross section to be zero below the $E_{^8Be}$ threshold, i.e., $\sigma_{\alpha\alpha}\left(E_{\alpha\alpha}<|Q_{\alpha\alpha}|\right)=0$ \cite{Depastas2024EPJ}, which of course is not applied to other elastic and inelastic channels.\\ 
\indent We note that same ideas are also implied by Nomoto et al. \cite{Nomoto1985}. There, the authors include a term $E_{^8Be}-E_{\alpha\alpha}$ in the extrapolation-based calculation of the widths and S-factors, essentially arguing that even in the direct channel, the energy missing from a pair with $E_{\alpha\alpha}<E_{^8Be}$, is given by the third $\alpha$-particle.\\
\indent The presence of the $E_{^8Be}$ threshold profoundly affects the determination of the major $\alpha-\alpha$ energy contribution. To demonstrate this, we consider a simple parametrization for the cross section as:
\begin{equationc}
\sigma_{\alpha\alpha}(E)=\frac{S_{\alpha\alpha}(E)\theta(E-E_{^8Be})}{Ee^{\xi_{\alpha\alpha}/\sqrt{E}}}
\label{e9a}
\end{equationc}
where $S_{\alpha\alpha}(E)$ is the astrophysical factor and $\xi_{\alpha\alpha}=\frac{Z_{\alpha}^2e^3\pi\sqrt{2\mu}}{\hbar}=5.60$ MeV$^{1/2}$. The energy that contributes most to the reaction rate, is then determined by the maximal value of the integrad of $\langle\sigma_{\alpha\alpha}v\rangle$, as given in Eq. \ref{e6a}. This is achieved via the solution of the equation
\begin{equationc}
\theta(E-E_{^8Be})\left[\frac{dlnS}{dE}+\frac{\xi_{\alpha\alpha}E^{-3/2}}{2}-\frac{1}{kT} \right]+\delta(E-E_{^8Be})=0
\label{e9b}
\end{equationc}
where we can distinguish three cases. For sub-threshold energies $E<E_{^8Be}$, $\theta(E-E_{^8Be})=\delta(E-E_{^8Be})=0$, so the solution is identical. For $E\rightarrow E_{^8Be}$, $\theta(E-E_{^8Be}) \rightarrow 1$ and $\delta(E-E_{^8Be})\rightarrow \infty$, which implies that $kT \rightarrow 0$. This is essentially equivalent to the collapse of the integrad into a $\delta$-function centered around $E_{^8Be}$, at low temperatures, i.e., $e^{-E/kT}\rightarrow\delta(E-E_{^8Be})$. This forces the $3\alpha$ system at the total energy of $\sim\frac{3}{2}E_{^8Be}$, without the requirement of an exact TS resonance. A lower bound for temperatures that show this behavior may be obtained by considering $E>E_{^8Be}$ solutions, which in the $\frac{dlnS}{dE}\sim 0$ approximation yield $kT>\frac{2E_{^8Be}^{3/2}}{\xi_{\alpha\alpha}}\sim 10$ keV. We stress that the simple parametrization of Eq. \ref{e9a} is used exclusive for the demonstration of the previous arguments and not for the calculations that follow. We specifically observe that a more accurate calculation of the integrad quantity via the H$\alpha$C model (data from Ref. \cite{Depastas2024EPJ}), show the $\delta$-function behavior at low temperatures, but set the lower bound at $kT\sim 5$ keV. This is presented in Fig. \ref{Fig5}, where $\frac{\sigma_{\alpha\alpha}(E)f_{MB}(E)}{\sigma_{\alpha\alpha}(E)f_{MB}(E)|_{E=95 keV}}$ is plotted as a function energy, for different temperatures according to the key.\\

\begin{figure}[!ht]                                        
\centering
\includegraphics[height=7.0 cm]
{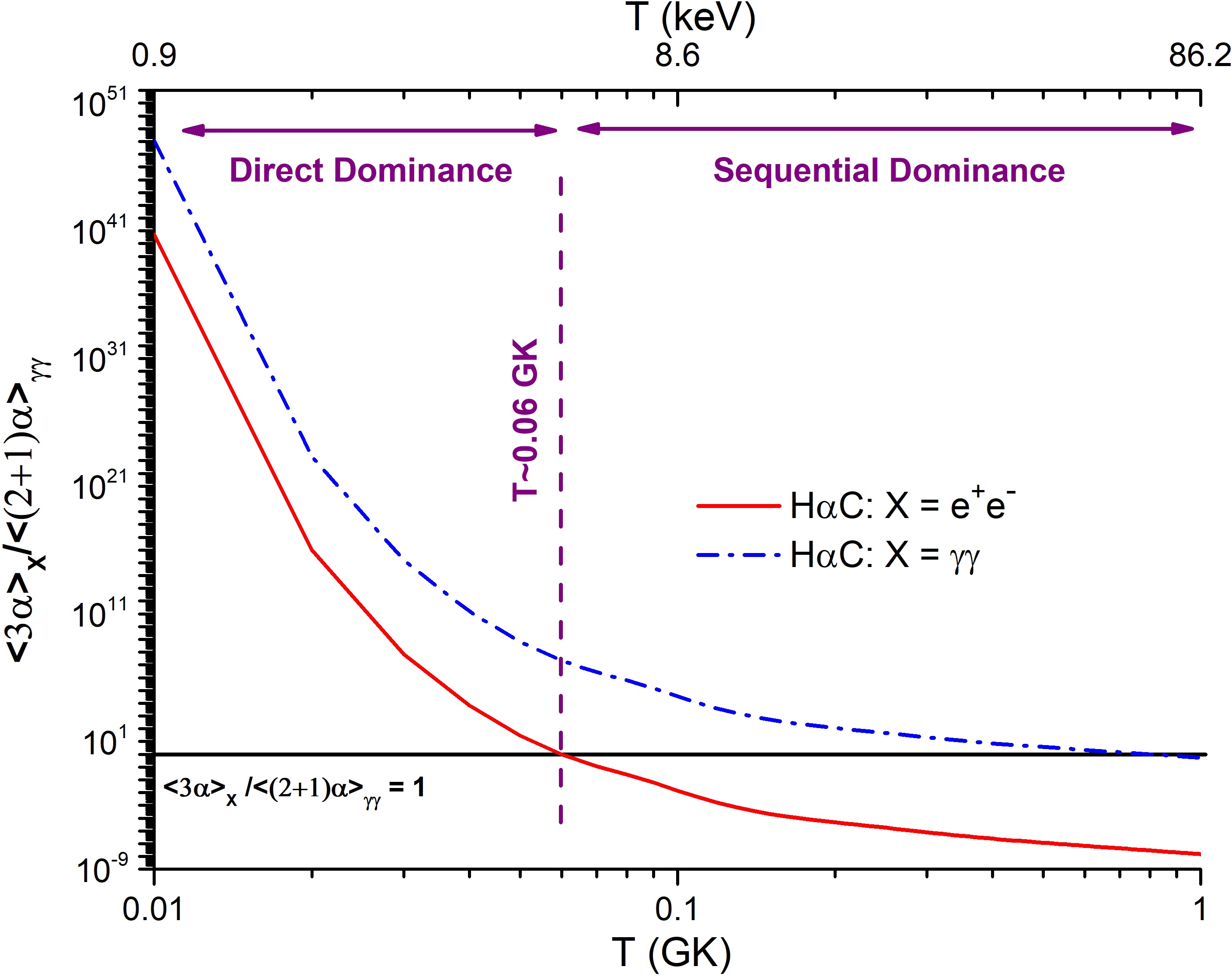} 
\caption{(Color online) The 3-body to (2+1)-body reduced reaction rate ratio as a function of stellar temperature. The numerator is calculated with either the $\gamma\gamma$ or $e^+e^-$ decays, according to the key.}
\label{Fig2}
\end{figure}

\begin{figure}[!ht]                                        
\centering
\includegraphics[height=12.5 cm]
{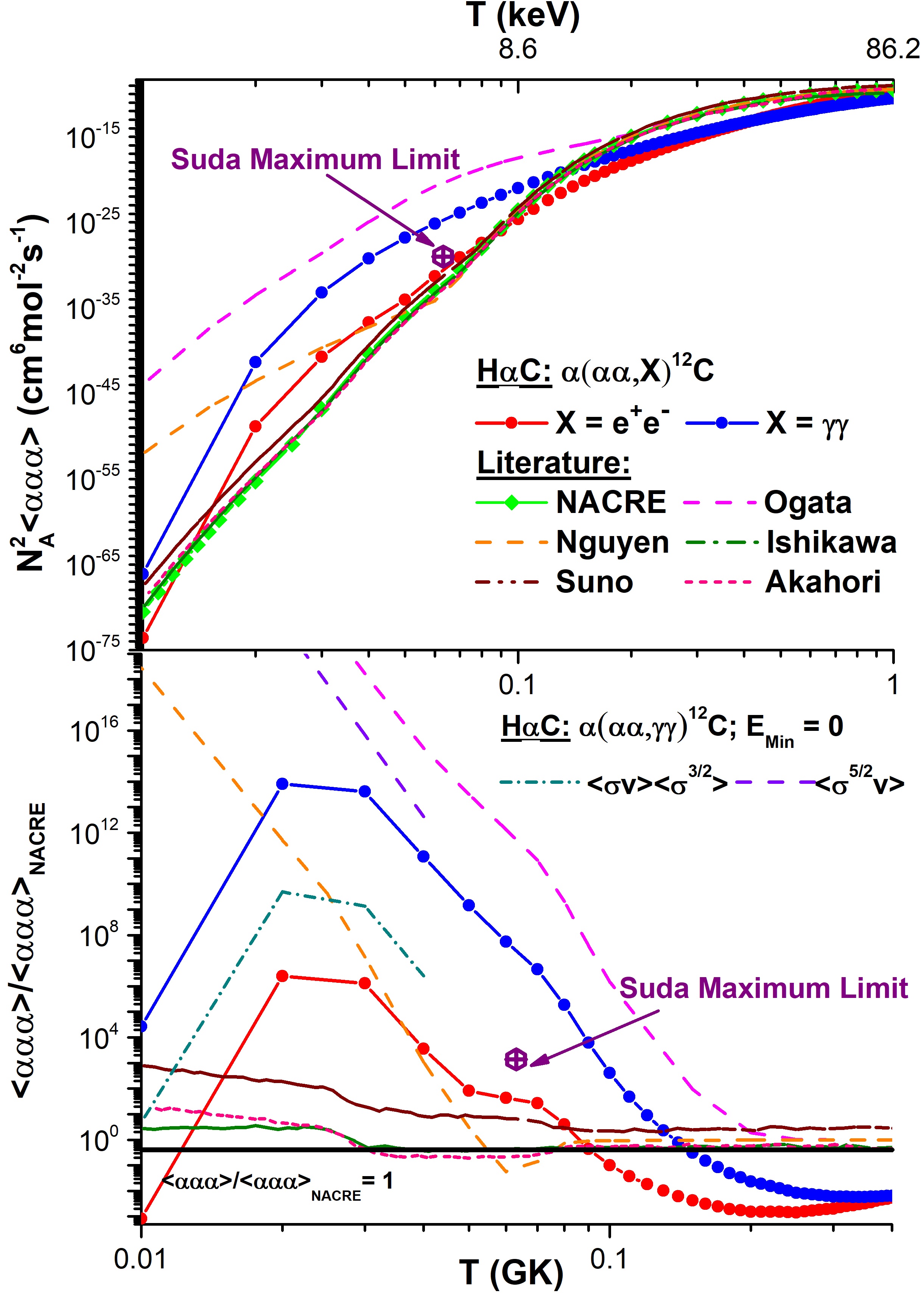} 
\caption{(Color online) Total reduced reaction rate $\langle \alpha\alpha\alpha\rangle=\langle \left(2+1\right)\alpha\rangle+\langle3\alpha\rangle$ as a function of stellar temperature (top) and the same quantity normalized by the NACRE data \cite{Angulo1999} (bottom). The results of $\gamma\gamma$ and $e^+e^-$ exit channels are presented along with theoretical data from Ref.s \cite{Angulo1999,Ogata2010,Nguyen2013,Ishikawa2013,Akahori2015,Suno2016}, according to the key. We furthermore signify the astrophysical upper limit by Suda \cite{Suda2011}. The additional $E_{min}=0$ results in the bottom panel are shown for comparison to the CDCC techniques, as explained in the text.}
\label{Fig3}
\end{figure}

\begin{figure}[!ht]                                        
\centering
\includegraphics[height=7.0 cm]
{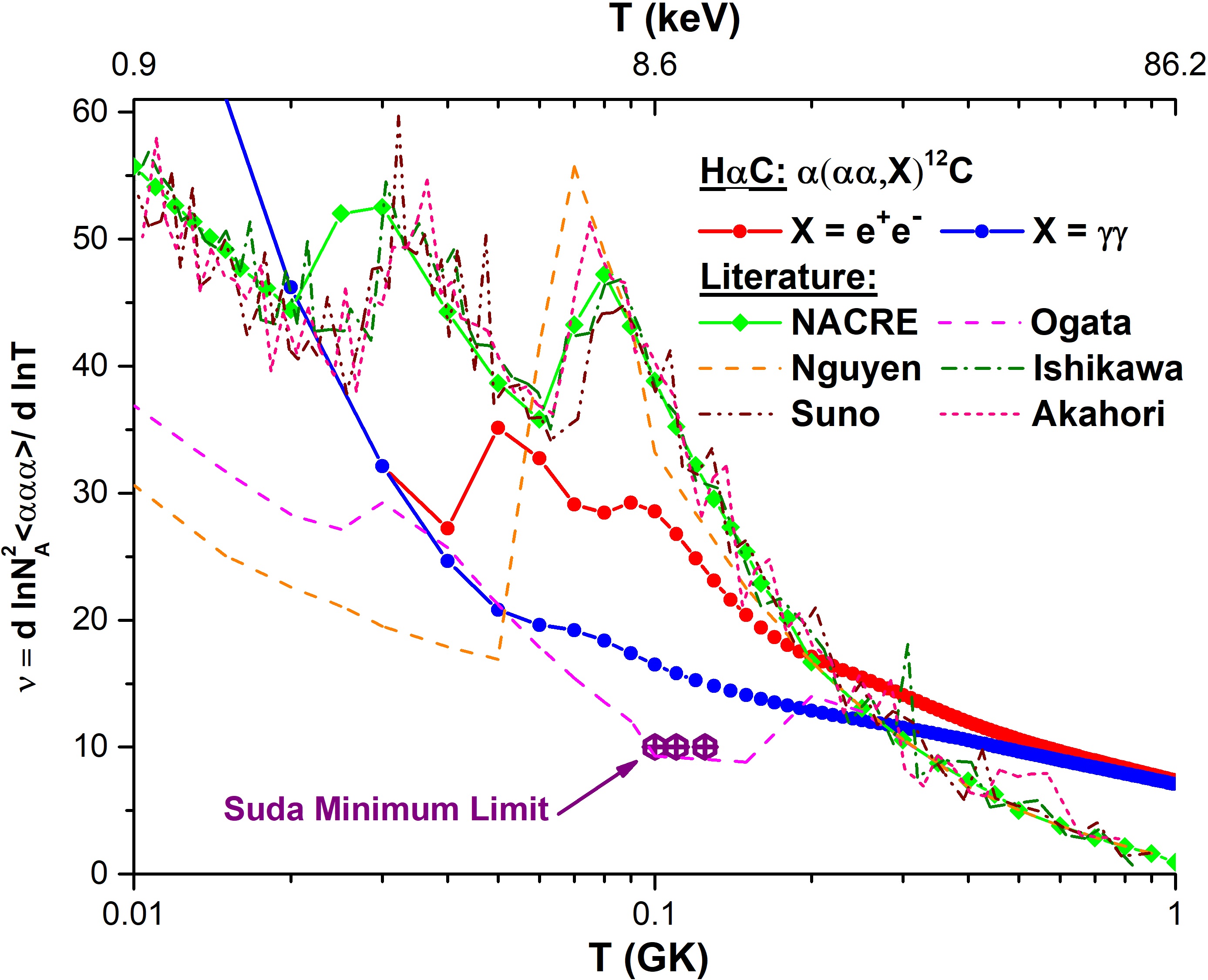} 
\caption{(Color online) Temperature dependence of the total reduced reaction rate as a function of stellar temperature. The results of $\gamma\gamma$ and $e^+e^-$ exit channels, as well as, theoretical data from Ref.s \cite{Ogata2010,Nguyen2013,Ishikawa2013,Akahori2015,Suno2016} and the lower Suda constraint \cite{Suda2011} are shown with similar notation to Fig. \ref{Fig3}.}
\label{Fig4}
\end{figure}

\indent Having discussed the fusion and kinetics of the sequential and direct mechanisms, we turn our attention to the exit channels. The consensus for the ES decay is via two successive $L=2$ photons, first to the $2_1^+$, $4.43$ MeV state and then to the ground state \cite{Caughlan1988,Nomoto1985,Angulo1999}. Even though the $\gamma\gamma$-decay scheme is also adopted for the direct channel in the current literature \cite{Nomoto1985,Ogata2010,Nguyen2013,Ishikawa2013,Akahori2015,Suno2016}, a close inspection of the compound nucleus symmetry properties prohibits it in our case (TS).\\
\indent When the direct mechanism dominates, the energies of each $\alpha-\alpha$ pair must be approximately equal, as is discussed before, while in any other case a sequential process would take place. This makes the geometry of the fused system an almost equilateral triangle, which is similar to the TS (although not exact since it has excitation energy) and termed as ``$0^+_E$". Its coupling with an $L=2$ photon then, would induce an angular momentum transfer in opposite direction of the photon's spin, This could only happen if two $\alpha$-particles were closer to each other, which is not the case with the TS.\\
\indent This argument can formally stated in terms of group theory. The equilateral triangle belongs the $D_3$ point group \cite{CottonGroupTheory}. Specifically for a coupling along the symmetry axis, the breathing mode of $0^+_E$ with respect to the hyperradius $R$ \cite{ZhengEfimov2020} and the quadrupole transition operator $\hat{Q}_{20}$ span the totally symmetric $A_1$, while the $2^+_1$ state spans the $E$ irreducible representation. The symmetry of the transition matrix element is then, $ \langle 2^+_1 | \hat{Q}_{20} | 0^+_E\rangle \sim E \otimes A_1 \otimes A_1 =E$ and since it does not contain the totally symmetric representation, the $E2$ transition is suppressed, i.e., $A_1 \notin \langle 2^+_1 | \hat{Q}_{20} | 0^+_E\rangle \sim E \Rightarrow \langle 2^+_1 | \hat{Q}_{20} | 0^+_E\rangle =0$.\\
\indent To avoid the symmetry concerns raised by angular momentum coupling, we are driven to examine the $E0$ decay path via the $e^+e^-$ pair production, which is one of the known decay modes of the Hoyle state \cite{LuoPRC2024,ChernykhPRL2010}. One possible approach would be the Schwinger mechanism \cite{Schwinger1951}. We could essentially follow Ref.s \cite{SettlemyreParticles2022,SettlemyrePRC2023,Depastas2024EPJ_PairProd} and assume a symmetric production of $e^+e^-$, but from a simple inspection of Eq. (7) of Ref. \cite{SettlemyreParticles2022}, we see that the electric field is not strong enough to produce an $e^+e^-$ pair for this sub-barrier regime. Thus, we adopt a perturbative approach similar to Wilkinson \cite{WilkinsonNPA1969}.\\
\indent The $E0$ decay probability per unit time $dP_{e^+e^-}/dt$ is factorized as \cite{WilkinsonNPA1969,ChernykhPRL2010}:
\begin{equationc}
\frac{dP_{e^+e^-}}{dt}=\frac{dP_{e^+e^-}^{Z=0}}{dt}C(Z=6)
\label{e10}
\end{equationc}
with $dP_{e^+e^-}^{Z=0}/dt$ and $C(Z=6)$ the nuclear and Coulomb contributions, respectively. The latter is proportional to $\left(E^*\right)^{1-\xi}$, where $\xi=\sqrt{1-\alpha^2Z^2}$ and $\alpha=1/137$ the fine structure constant. For $Z=6 \rightarrow\xi\approx0.999$, so the Coulomb correction is approximately taken as a constant, evaluated at the Hoyle state, i.e., $C(Z=6)\approx C_H=1.1013$, which is the value calculated for the Hoyle state \cite{ChernykhPRL2010}. The nuclear contribution is given by \cite{WilkinsonNPA1969}:
\begin{equationc}
\frac{dP_{e^+e^-}^{Z=0}}{dt}=\frac{8e^4m_e^5c^4}{27\pi\hbar^7}|M|^2\left(\frac{E^*}{2m_ec^2}-1\right)^3\left(\frac{E^*}{2m_ec^2}+1\right)^2B(S)
\label{e11}
\end{equationc}
where $E^*$ is the excitation energy of the compound nucleus, $m_e$ the mass of the electron/positron, $|M|$ the modulus of the nuclear matrix element and $B(S)$ phase-space correction, with $S=\left(E^*-2m_ec^2\right)/\left(E^*+2m_ec^2\right)$. For the latter we use the 5$^{th}$ order polynomial expansion in terms of $S$ \cite{WilkinsonNPA1969}:
\begin{equationp}
B(S)\approx\frac{3\pi}{8}\left(1-\frac{S}{4}-\frac{S^2}{8}+\frac{S^3}{16}-\frac{S^4}{64}+\frac{5S^5}{512}\right)
\label{e12}
\end{equationp}
\indent The nuclear matrix element for the transition is defined as $M\equiv\langle 0^+_1 | \sum_p r_p^2 | 0^+_E\rangle$ \cite{OppenheimerPR1939}, where the sum is understood to run over all the protons. An accurate estimation of this would require a quantum mechanical calculation, such as an expansion over the shell model states. Here, we obtain an approximate form for low excitation energies, based on physical intuition. We expand the identity over all the $J^{\pi}$ excited nuclear states:
\begin{equationp}
M=\langle 0^+_1 | \sum_p r_p^2 | 0^+_E\rangle = \sum_{J^{\pi}} \langle 0^+_1 | \sum_p r_p^2 | J^{\pi} \rangle  \langle J^{\pi} | 0^+_E\rangle
\label{e13}
\end{equationp}
\indent Taking orthogonality and parity conservation into account, the overlaps are $\langle J^{\pi} | 0^+_E\rangle\sim \delta_{J,0}\delta_{\pi,+}$. The matrix element then is given only by the $0^+$ states:
\begin{equationp}
M=\sum_{N\ge 1} \langle 0^+_1 | \sum_p r_p^2 | 0^+_N \rangle  \langle 0^+_N | 0^+_E\rangle
\label{e14}
\end{equationp}
\indent The major contributions to the sum correspond to high values of overlap integrals $\langle 0^+_N | 0^+_E\rangle$, with states similar in energy and geometry. This rough approximation is based on the low-energy character of the $3\alpha$ reaction and is widely used for the vibronic coupling of molecular states, where it is known as the Frank-Condon principle \cite{HollasSpectrosocpy}. Consequently, the matrix element is mostly determined by the Hoyle state coupling, as the ground state is not clusterized and the $0^+_{N>2}$ states are above the Coulomb barrier, i.e., $\langle 0^+_2 | 0^+_E\rangle \gg \langle 0^+_{N\ne 2} | 0^+_E\rangle$. We furthermore expect, that the value of the matrix element reduces with $|E_H-E|$ and as such, we parametrize $|M|^2$ with a Gaussian form:
\begin{equationc}
|M|^2\approx\left|\langle 0^+_1 | \sum_p r_p^2 | 0^+_2 \rangle  \langle 0^+_2 | 0^+_E\rangle\right|^2\equiv \langle r^2 \rangle_{tr}e^{-(E-E_H)^2/b^2}
\label{e15}
\end{equationc}
where $\langle r^2 \rangle_{tr}$ is the Hoyle state matrix element, experimentally extracted at $5.29$ fm$^2$ \cite{ChernykhPRL2010} and $b$ a fitting parameter. This is evaluated at $b=1.732$ MeV, by applying a $1\%$ cut-off to the coupling probability of the $0^+_3$, $9.930$ MeV state \cite{NNDC2022}.\\
\indent The final relation of the $e^+e^-$ probability is derived by multiplication of the the decay rate (Eq. \ref{e10}) with the characteristic time interval $\delta \tau_E$ of the $O^+_E$ breathing motion. This is calculated by integrating the relative kinetic energy of 3 $\alpha$-particles at the equilateral positions \cite{ZhengEfimov2020}, between the turning points of the potential well $V_{\alpha\alpha\alpha}(R)$, $R_N$ and $R_{in}$:
\begin{equationc}
\delta \tau_E = \int_{R_N}^{R_{in}} \frac{dR}{\dot{R}}=\int_{R_N}^{R_{in}} dR \sqrt{\frac{m_{\alpha}}{2\left[E_{\alpha\alpha\alpha}-V_{\alpha\alpha\alpha}(R)\right]}}
\label{e16}
\end{equationc}
where $m_{\alpha}$ is the mass of an $\alpha$ particle, $V_{\alpha\alpha\alpha}(R)=3V_{\alpha\alpha}(R)$, $V_{\alpha\alpha}$ the $\alpha-\alpha$ potential of the H$\alpha$C model \cite{ZhengHac2021,Depastas2023} and $E_{\alpha\alpha\alpha}$ the 3-body center of mass energy. The equal energies requirement for a Thomas mechanism is enforced through the relation $E_{\alpha\alpha\alpha}=\frac{3}{2}E_{\alpha\alpha}=E^*-Q_{\alpha\alpha\alpha}$, with $Q_{\alpha\alpha\alpha}=7.2746$ MeV \cite{Angulo1999}. Combining Eq.s \ref{e10}-\ref{e16}, the decay probability is written as:
\begin{equationp}
P_{e^+e^-}\approx\frac{dP_{e^+e^-}^{Z=0}}{dt}\left(E_{\alpha\alpha}\right)\delta \tau_E\left(E_{\alpha\alpha}\right)C_H
\label{e17}
\end{equationp}
\indent The error of our approach to the $e^+e^-$ decay probability for energies $E^*/m_ec^2\sim 15-20$, is consisted of the $\sim1\%$ error of the Coulomb factor (Fig. 2 of Ref. \cite{WilkinsonNPA1969}), the $\sim0.5\%$ error of the phase-space correction and the $\sim1\%$ in the matrix element couplings. In total this results in less than $2\%$ uncertainty, which is acceptable considering the large reaction variations in the literature.\\
\indent To further demonstrate that the $E2$ exit path is not appropriate for the direct channel, even if the symmetry considerations are neglected, we calculate the $P_{\gamma\gamma}$ probability adopting our sequential framework (Ref. \cite{Depastas2024Plb}, Eq.s 5, 7 and 10). The tunneling time and penetrability are fitted to the H$\alpha$C data \cite{Depastas2024EPJ}, according to:
\begin{equationc}
\tau_{\alpha}(E)\sim \frac{617.54648}{E^{1.51418}}\text{ fm/c}
\label{e18}
\end{equationc}
\begin{equationc}
T_{0}(E)=\frac{1}{1+e^{2A/\hbar}};\frac{A}{A_G}\sim 0.09887+0.75783e^{-1.99941E}
\label{e19}
\end{equationc}
where $A$ and $A_G$ are the total and Gamow sub-barrier actions \cite{Depastas2023}, respectively, while $E= E_{\alpha\alpha}/2<0.5$ MeV, in accordance with the geometry of the problem.\\
\indent The ratio of our results for the direct $\langle3\alpha\rangle_X$ (via both $X=e^+e^-$ and $X=\gamma\gamma$ decays) and sequential $\langle\left(2+1\right)\alpha\rangle_{\gamma\gamma}$ (which decays always via the $\gamma\gamma$ channel) reduced reactions rates (Eq.s \ref{e7} and \ref{e8}) as functions of stellar temperature is presented in Fig. \ref{Fig2}. We observe that with the $E0$ decay, the direct channel is dominant for $T\lesssim0.06$ GK, which agrees with the prediction of Ref. \cite{Nomoto1985}, while with the $E2$ it is dominant for the entire $T=0.01$ GK$-1$ GK region. This contradicts the predictions of all the available literature and further signifies its inappropriateness as a possible exit channel.\\
\indent The total reaction rate $\langle \alpha\alpha\alpha \rangle$ is taken as the sum of the direct and sequential channel contributions. Our results are shown in the top panel of Fig. \ref{Fig3}, along with several available theoretical data sets \cite{Angulo1999,Ogata2010,Nguyen2013,Ishikawa2013,Akahori2015,Suno2016}, while the corresponding curves normalized by the NACRE results \cite{Angulo1999} are depicted in the bottom panel.\\
\indent Firstly, we observe that the $\alpha(\alpha\alpha,\gamma\gamma)^{12}$C reaction rate is $8-10$ orders of magnitudes higher than the $\alpha(\alpha\alpha,e^+e^-)^{12}$C rate for temperatures $T<0.2$ GK, after which almost all the methods are in agreement. Furthermore, there is a proximity between the $E2$ and Continuum-Discretized Coupled Channels (CDCC) results of Ogata et al. \cite{Ogata2010}. Both are excluded by the maximum astrophysical upper limit set by Suda et al. \cite{Suda2011}, while their difference at low temperatures comes from the choice minimum integration energy. This is shown in the bottom panel of Fig. \ref{Fig3}, where we also include the $\alpha(\alpha\alpha,\gamma\gamma)^{12}$C result (purple dashed curve) integrated from $E_{min}=0$ energy.\\
\indent The results of Ref.s \cite{Nguyen2013,Ishikawa2013,Akahori2015,Suno2016} on the other hand, are at most $3$ orders of magnitude higher than the NACRE data at low temperatures. This is attributed to the level of truncation of the $\alpha+\alpha$ CDCC states \cite{Akahori2015}. We note that all of these approaches contain an attractive genuine 3-body force, to fit the $0_1^+$ and $2_1^+$ states, simultaneously with the $0^+_2$. This technique allows the integration from zero energy and decreases the Coulomb barrier which, counter-intuitively hinders the penetrability for channels above it \cite{Hagino2022}. Thus, the larger the model space contains more suppressed above-the-barrier channels which reduce the rates closer to NACRE. Nevertheless, the attractive character of the 3-body force leads to the collapse of an extended $\alpha$-conjugate system and therefore, is unphysical.\\
\indent In our approach, this essentially corresponds to a separation of the successive scatterings with different averaging integrals for each and the latter taking into the decay probability, i.e., Eq. \ref{e7} is rewritten as $\langle3\alpha\rangle_{\gamma\gamma} \rightarrow\frac{4}{3\sqrt{\pi}}\langle\sigma_{\alpha\alpha}v\rangle \langle \sigma_{\alpha\alpha}^{3/2}v P_{\gamma\gamma}\rangle$ and integrated from $E_{min}=0$. This approximation is presented in the bottom panel of Fig. \ref{Fig3} (cyan dash-dot line) and its proximity to the results of Ref.s \cite{Nguyen2013,Ishikawa2013,Akahori2015} at $T=0.01$ GK is striking. The three $\alpha$-particles are strongly correlated though far below the Coulomb barrier. These correlations are the essence of the Thomas-Efimov theorem and as such, this splitting is not an accurate physical description, even if it obeys the astrophysical constraints.\\
\indent The $\langle3\alpha\rangle_{e^+e^-}+\langle\left(2+1\right)\alpha\rangle_{\gamma\gamma}$ results of the $\alpha(\alpha\alpha,e^+e^-)^{12}$C oscillate around the NACRE values (black line in bottom panel of Fig \ref{Fig3}). It yields lower rates by 3 orders of magnitude at $T=0.01$ GK and higher at $T=0.02-0.1$ GK, peaking at an $10^6$ enhancement. It remains at least 3 orders of magnitude lower than the Suda astrophysical limit \cite{Suda2011}, while also obeying the recent stringer constraint based on an R-Matrix analysis \cite{Santra2025} at the same temperature. We note that our choice of the H$\alpha$C data for the $\alpha-\alpha$ fusion cross section which do not show a resonant enhancement at $E_{^8Be}$, instead of the NM data which they do, is important. As shown in Ref. \cite{SunPRC2025}, the $\alpha-\alpha$ fusion cross section is increased by a factor of $\sim 25$, which in turn gives higher rates. This effectively includes the predicted TS at $7.458$ MeV \cite{Cardella2022,ZhengEfimov2020,ZhengHoyle2018,BishopEfimov2021,BaishyaEfimov2025}, which has not been observed yet and would require a re-evaluation of the astrophysical and R-Matrix constraints.\\
\indent The final test for our method is the calculation of the temperature dependence $\nu$ of the reaction rates. This is presented in Fig. \ref{Fig4}, in comparison with the corresponding literature values. The notation is the same as in Fig. \ref{Fig3}. Our results respect the $\nu\gtrsim10$ minimum astrophysical limit by Suda et al. \cite{Suda2011}, which ensures He-flashes in AGB stars. Although this trial excludes the direct frameworks of Ogata et al. \cite{Ogata2010} and Garrido et al. \cite{Garrido2011} (not shown in Fig. \ref{Fig4}), it cannot distinguish between the $\gamma\gamma$ and $e^+e^-$ exit channels of our approach.\\
\indent To conclude, here we study the $3\alpha$ process through the lenses of a unified Thomas-Efimov theory. We recognize the sequential channel as an Efimov based and the direct as a Thomas based mechanism. As such, we expand our previous theoretical framework describing the sequential channel, to accommodate for the direct channel, which is known to dominate the low temperatures region. A novel approach for the simultaneous N-body scattering is developed based on successive 2-body collisions, that effectively exchange 1 particle between the rest $N-1$. The advantages of this methodology is the simplification of long range correlations for $N>2$ charged particle scattering, as well as its general many-body character. Its application to He-burning is achieved utilizing the general formulas for $N=3$ with the $\alpha$-$\alpha$ fusion cross sections in imaginary time.\\
\indent For the exit channel, we consider both the $E0$ and $E2$ possible paths. We demonstrate in terms of both group theory and physical intuition, the latter is prohibited. The $e^+e^-$ decay is approached by adopting a perturbative scheme in the equilateral (i.e., TS) geometry. The major contribution to our approach is calculated with an estimated maximum error of $2\%$.\\
\indent Our favor towards the $e^+e^-$ exit path is augmented by the calculation of the total reduced reaction rates. While the $\alpha(\alpha\alpha,e^+e^-)^{12}$C rates confirm our expectations of low temperature dominance and obey both the astrophysical \cite{Suda2011} and R-matrix based \cite{Santra2025} constraints, the $\alpha(\alpha\alpha,\gamma\gamma)^{12}$C reaction proceeds many orders of magnitude faster. The result oscillates around the NACRE \cite{Angulo1999} data and provides a sound description of the strongly correlated low temperature region, both in its nuclear, astrophysical and group theoretical aspects.
\\ACKNOWLEDGMENTS\\
This work work was supported in part by the United States Department of Energy under Grant $\#$DE-FG02-93ER40773.
\bibliographystyle{model1-num-names}
\bibliography{references}{}
\end{document}